\documentclass[12pt,aps,preprint]{JHEP3}
\usepackage{amsmath,amssymb,graphicx,mathrsfs,epsfig}
\usepackage[latin1]{inputenc}
\usepackage{bm}
\newcommand{\be}{\begin{eqnarray}}
\newcommand{\ee}{\end{eqnarray}}
\renewcommand{\H}{{\cal H}}

\title{Expressing the equation of
state parameter \\ in terms of the three dimensional cosmic shear}
\author{Daniel Levy, Ram Brustein\\
Department of Physics, Ben-Gurion University,
    Beer-Sheva 84105, Israel \\ E-mail: levyda@bgu.ac.il,ramyb@bgu.ac.il}
\preprint{}
\abstract{
We study the functional dependence of the spin-weighted angular moments of the two-point correlation function of
the three dimensional cosmic shear on the expansion history of the universe.
We first express the redshift dependent total equation of state parameter in terms of the growing mode of the gauge invariant metric perturbation in the conformal-Newtonian gauge for the case of adiabatic perturbations. We then express the redshift dependent angular moments of the shear two-point correlation function as an integral in terms of the metric perturbation. We present
the final explicit expression for the case of a Harrison-Zeldovich spectrum of
primordial perturbations. Our analysis is restricted to the linear regime.  We use our results to make a preliminary study of the required sensitivity that will allow cosmic shear observations to add significant information about the expansion history of the universe.
}
\begin{document}

\section{Introduction}

Determining the expansion history of the universe is one of the central problems in
cosmology, and the goal of many observation programs: distant supernovae
\cite{sn1,sn2,sn3}, the large scale structure in the universe \cite{sdss1,sdss2,Tegmark:2006az}, and
the cosmic microwave background \cite{Komatsu:2008hk}. It has become clear in recent years
that kinematic distance probes in the homogeneous and isotropic universe, such as
luminosity distance, angular distance etc., have a limited resolving power for
determining the expansion history of the universe. They rely on light emitted from
distant sources, and hence they measure an integral over the expansion history
\cite{maor1,weller,Barger:2000jg}. It is possible to improve the determination of
the expansion history by adding prior assumptions on the evolution, or by combining
several kinds of observations \cite{maor2,Frieman:2002wi}, or by focusing on better
determined quantities \cite{Corasaniti:2002vg,Alam:2003sc,Wang:2004ru}. It has also
been suggested that the measured integrals can be differentiated
\cite{Daly:2004gf}.  Reviews of the subject can be found in \cite{copeland} and \cite{Frieman:2008sn}.

Cosmological perturbations provide, through their dependence on the homogeneous and
isotropic background,  an independent tool to probe the expansion history of the
universe \cite{Linder:2003dr,Cooray:2003hd,Linder:2005in}. To be able to use the
perturbations to determine the expansion history of the universe it is necessary to
express the total equation of state parameter $w_{\rm tot}$ in terms of the
perturbations in a way that does not depend on the specific functional dependence
of $w_{\rm tot}$ on time.  In the following we refer to such an expression as
``model-independent".  Further, it is
necessary to identify observable quantities from which one can determine in a
reliable way the perturbations as a function of time, or equivalently, of redshift.
Obviously, we have to look for observables that can be measured precisely, however,
in addition they also have to be evaluated precisely, otherwise the theoretical
errors form the imprecise calculation will dominate the final error budget and will
limit their resolving power as probes of the expansion history of the universe.

The three dimensional cosmic shear \cite{3dwl1,3dwl2,3dwl3} seems to be a promising
observable that can be measured by weak lensing observations. Currently, the
observations are mostly two dimensional \cite{wlobs1,wlobs2,wlobs3} with preliminary studies of 3D analysis \cite{Kitching:2006mq,Massey:2007gh} and future programs expected to have three dimensional capabilities \cite{wlobs4,Amara:2006kp}.
The recent advances in measuring the cosmic shear and the expected future
improvements have attracted a growing interest in the potential of weak lensing
measurements for determining the expansion history of the universe, either by
itself \cite{Huterer:2001yu,Hu:2002rm,Jain:2003tb,Ishak:2004jj,Simpson:2004rz} or
in combination with other measurements
\cite{Upadhye:2004hh,Ishak:2005we,Schimd:2006pa}. Cosmic shear is a measure of the
shape distortion of  distant objects  as the light emitted from them propagates
through the perturbed universe.  To measure the three dimensional cosmic shear the
redshift of the distorted objects (usually galaxies) needs to be measured in
addition to the distortion pattern. Several recent reviews of weak lensing give a
comprehensive and exhaustive description of the current state of the field
\cite{wlrev1,wlrev2,wlrev3,wlrev4}.

If the perturbations are weak, the cosmic shear depends linearly on them. In this
linear regime of the metric perturbations, both observations and theoretical
calculations can be done reliably without any obvious obstruction.  Thus, it seems
possible to reduce the intrinsic theoretical and experimental errors down to the
percent level, as required for accurate determination of the expansion history of
the universe. In gauge invariant perturbation theory, the metric perturbation
$\Phi$ can in principle be small even if the density perturbations $\delta
\rho/\rho $ are not, thus the use of metric perturbations allows us to extend the
linear regime to smaller scales.

The standard approach of exploring the relationship between the cosmic shear and the expansion history is to use numerical methods. Numerical comparison to cosmological models is made and used to estimate the prospects of constraining cosmological parameters from shear measurements. This method gives precise results for specific models and for specific parametrizations.  In this approach the main important factors that need to be determined are related to the accuracy of the shear measurements and to the estimators for the power spectrum \cite{Schneider:2002jd}.

We wish to take a different approach. We pose a different question which we believe is quite significant and moreover can be answered in a definite way. The question that we wish to pose and answer is: What is the theoretical limit on model-independent information about the expansion history of the universe that can be obtained from cosmic shear measurements with a given accuracy? We wish to know, in theory, whether the 3D cosmic shear (or its angular power spectrum) is more sensitive to changes in the total equation of state parameter than other observables. Our results allow us to estimate the accuracy goal needed for shear measurements so they can improve on other accurate tests such as luminosity distance measurements or CMB measurements. Our approach is mostly relevant when trying to estimate the prospects of future lensing surveys for constraining the evolution of the universe in a model independent way.
To answer the question that we have just posed we need to determine the {\em functional dependence} of the cosmic shear on the expansion history of the universe. In the context of this paper this is equivalent to determining the  sensitivity of the angular spectrum of the 3D cosmic shear to changes in the total equation of state parameter.

 In this paper we find a model independent analytical solution for the growth of the metric perturbations (Section~\ref{a}) and show that the spatial and time dependent parts are separable under certain conditions. We  then use this solution to show the functional dependence of the three dimensional cosmic shear on the DE equation of state parameter (Section~\ref{b}) in the linear regime. We then explain how to extract  $w_{tot}$  from the red-shift evolution of the shear angular multipoles (Section~\ref{c}). Using our results we make a preliminary  analysis of the theoretical sensitivity to changes in the evolution of dark energy and estimate the influence of such changes on shape and strength of the shear angular spectrum. Section~\ref{e} contains our conclusions. In appendix~\ref{App1} we present a detailed derivation of the functional dependence of  the total equation of state parameter on the metric perturbation.

\section{Expressing $\Phi$ in terms of $w_{\rm tot}$}
\label{a}
\subsection{The perturbation equations}

The line element of the perturbed universe in the conformal Newtonian
(longitudinal) gauge is
\begin{equation}
 \label{pertle}
ds^2=a^2(\eta)\left[-(1+\Phi\left(\eta,\vec{x})\right)d\eta^2
+\left(1-\Phi\left(\eta,\vec{x}\right)\right)\left(dw^2+f_K^2(w)d\Omega^2\right)\right].
\end{equation}
$\eta$ is the conformal time,  $d\Omega$ is the two sphere differential element,
and $f_K$ depends on the spacial curvature $K$,
\begin{eqnarray}
\label{fk}
    f_K(w)=\left\{\begin{array}{ccc}
                K^{-1/2}\sin (K^{1/2}w) & , & K>0 \\
                w & , & K=0  \\
                (-K)^{-1/2}\sinh ((-K)^{1/2}w) & , & K<0  \\
               \end{array} \right..
\end{eqnarray}
We choose to ignore shear perturbations and assume that the cosmic fluid is a
perfect fluid. Imperfect fluid perturbations' influence is in general considered to
be small  (see for example \cite{Koivisto:2005mm}) and could be evaluated in
subsequent investigations. We follow the standard derivations that are reviewed in
\cite{mukhanov} to obtain the equations of motion for $\Phi$. They are derived from
the perturbed Einstein's equations,
\begin{eqnarray}
 \label{deltarhoeq}
\nabla^2\Phi-3{\H} \Phi'-3 (\H^2-K)\Phi &=& 4\pi |K|  G a^2 \delta \rho, \\
 \label{deltapeq}
 \Phi''+3 \H \Phi'+ \left(2 \H'+(\H^2-K)\right)\Phi &=& 4\pi |K|
G a^2 \delta p.
\end{eqnarray}
where $\H= a'/a$, and prime denotes a derivative with respect to $\eta$.
The limit $K\to 0$ of these equations has to be taken carefully keeping $K a^2$
fixed,

As mentioned above, the linear equations are valid when $\Phi$ is small $|\Phi|\ll
1$, however this does not require in general that $\delta \rho/\rho$ is small. From
eq.(\ref{deltarhoeq}) we can estimate that for small wavelengths $\delta \rho/\rho$
is larger than $\Phi$ by a factor of the order of the square of the ratio of the
size of the horizon to the wavelength, $\delta \rho/\rho \sim (q/\H)^2 \Phi$.

Substituting $\delta p = (\frac{\partial \delta p}{\partial \delta\rho})_S \delta
\rho +(\frac{\partial \delta p}{\partial S})_\rho \delta S= c_S^2\delta \rho+\tau
\delta S $,  where the total speed of sound of the perturbations is $c_S^2=
\frac{\partial \delta p} {\partial \delta\rho}$ and $\delta S$ is the total entropy
perturbation, leads to a single second order equation for $\Phi$,
\begin{equation}
\Phi''+3 \H (1+c_S^2) \Phi'-c_S^2 \nabla^2\Phi+ (2 \H'+(1+3c_S^2)(\H^2-K))\Phi=4\pi
 |K|G a^2\tau \delta S.
\end{equation}
In the rest of the paper we will only consider adiabatic perturbations for which
$\delta S=0$. Then,
\begin{equation}
 \label{homperteq}
\Phi''+3 \H (1+c_S^2) \Phi'-c_S^2 \nabla^2\Phi+ \left[2
\H'+(1+3c_S^2)(\H^2-K)\right]\Phi=0.
\end{equation}
Here we can take the $K\to 0$ limit in a straightforward manner.

Our derivation is fully relativistic. In doing so we can put initial conditions on $\Phi$ outside
the horizon and follow its evolution, and hence use directly the early-time
information about the spectrum of metric perturbations from the CMB or the linear
matter power spectrum rather than using the late time processed matter power
spectrum. The numerical difference between the relativistic and non-relativistic
analysis at small redshift $z$ for small wavelength perturbations is not expected
to be large. Again, we can roughly estimate the difference from
eq.(\ref{deltarhoeq}) to be of the order of the square of the ratio of the size of
the horizon to the wavelength $\sim (q/\H)^2$.

In the perturbed Einstein equations (\ref{deltarhoeq}-\ref{deltapeq}) $\Phi$ is
related to the total density perturbation $\delta\rho$. In a model of a universe
containing dark energy and cold matter the solution for $\Phi$ depends on the two
component background and on both perturbations, the dark energy perturbation and
the cold matter perturbation. Equation (\ref{homperteq}) is always correct when the
total speed of sound  ${c_S}^2 =\frac{\partial \delta p_{\rm tot}}{\partial \delta
\rho_{\rm tot}}$ is used. If the various components are weakly coupled as expected for
matter and DE, then
\begin{equation}
    \frac{\partial \delta p_i}{\partial \delta \rho_j}=\delta_{ij}({c_S}_i)^2,
\end{equation}
and thus the total speed of sound is $ {({c_S})}^2 \delta \rho_{\rm tot} =\sum_i
{({c_S}_i)}^2 \delta \rho_i $. For the two fluid model
\begin{equation}
{({c_S})}^2 \delta \rho_{\rm tot} = {({c_S}_m)}^2 \delta \rho_m + {({c_S}_{DE})}^2
\delta \rho_{DE} .
\end{equation}
Here the subscript $m$ denotes matter quantities and the subscript $DE$ denotes DE
quantities.  The matter speed of sound vanishes $({c_S}_m)^2=0$, thus, for
adiabatic perturbations we get
\begin{eqnarray}
\label{twofluids}
 && \Phi''+3\left(1+\frac{\delta\rho_{DE}}{\delta\rho_{\rm
tot}}{({c_S}_{DE})}^2\right){\mathcal{H}}\Phi' -\frac{\delta\rho_{DE}}
{\delta\rho_{\rm tot}}{({c_S}_{DE})}^2\nabla^2\Phi
 \nonumber \\ &&
+\left[2{\mathcal{H}}'+ \left(1+3\frac{\delta\rho_{DE}}{\delta\rho_{\rm
tot}}{({c_S}_{DE})}^2\right)({\mathcal{H}}^2-K)\right]\Phi
    =0.
\end{eqnarray}

\subsection{The solution of the perturbation equations}

The general solution of eq.(\ref{homperteq}) is conveniently obtained by a standard
change of variables to $u$,
\begin{equation}
 \label{phiu}
\Phi=4 \pi G \sqrt{\rho+p}\ u.
\end{equation}
The equation for $u$ is
\begin{align}
 \label{uperteq}
 u''-c_S^2 \nabla^2 u - \frac{\theta''}{\theta} u & = 0, \\
\noalign{\noindent\text{with}} \cr
 \theta(\eta) =
 \frac{1}{a} \sqrt{\frac{\rho}{\rho+p}}& .
\end{align}
Here we assume spatial flatness. The expression and solutions for curved space can be found in \cite{mukhanov}.
Since the in the universe space curvature is known to be quite small we expect to be able to treat it as a perturbation in subsequent analysis.

We can express $\theta$ in terms of the time-dependent total equation of state
parameter
\begin{equation}
 \label{wtot}
w_{\rm tot}=\frac{p}{\rho},
\end{equation}
\begin{eqnarray}
 \label{thetaweq}
 \theta(\eta)=\frac{1}{a(\eta)} \frac{1}{\sqrt{1+w_{\rm tot}(\eta)}}.
\end{eqnarray}

Finding the full solution of the perturbation equations requires solving the two fluid equations. However, we are
interested in the growing solution at rather late times (say $z \lesssim 4$), when
substantial deviations from matter domination start to build up and for wavelengths
that are smaller than the horizon. At those late times the perturbations in the
cold matter will be the dominant perturbations and we will be able to safely ignore
the DE perturbation while taking into account the changes in the background
evolution due to DE (see \cite{Bean:2003fb,Hannestad:2005ak} for a recent
discussion).
As we explain below, if the perturbations are adiabatic then the initial amplitudes of the
different types of perturbations are about equal at horizon entry. During matter domination (when $w_{\rm tot}\approx 0$) the well known ``growing mode" solution of the perturbation equation is constant. DE perturbations, on the other hand, decay during matter domination.

To understand our argument more precisely, let us consider the following situation. Let us assume, for the moment,
that the DE perturbation dominates. The general  solution assuming that the DE perturbation
is the dominant one in Fourier space, for a mode with wave vector $\vec{q}$, is
\begin{eqnarray}
    u=\sqrt{\eta}(C_1 J_{5/2}(\sqrt{{c_S}^2 q}\eta)+C_2 Y_{5/2}(\sqrt{{c_S}^2 q}\eta)).
\end{eqnarray}
For large $\eta$ (late times) the Bessel functions decay as $1/\sqrt{\eta}$.
Consequently, the solution for $u$ approaches a constant at late times. From
eq.(\ref{phiu}), since during matter domination $\rho \sim 1/a^3$ and $a(\eta) \sim
\eta^2$, it follows that the solution for $\Phi$ decays as $\eta^{-3}$
\begin{eqnarray}
    \Phi \sim \eta^{-3}u(\eta) \sim   \eta^{-3}.
\end{eqnarray}
We see that if the DE and matter perturbations start off with equal
amplitudes, the DE perturbation will decay through the era of matter domination
with respect to the matter perturbation by a factor of
$\eta^{-3}=a^{3/2}=(1+z)^{-3/2}$. For example,  at $z=1$ a DE perturbation that
entered the horizon at $z=100$ will be smaller by a factor of about $10^{-3}$  than
a matter perturbation that entered the horizon at the same time with the same
amplitude.

Now, let us focus on the matter perturbations. We can solve the equation for the
matter perturbations in (\ref{twofluids}) with an arbitrary background equation of
state $w_{\rm tot}$. As the relative part $\frac{\delta\rho_{DE}}{\delta\rho_{\rm tot}}$ of the dark energy goes to zero the value of the total speed of sound $c_S^2 $ goes to zero. This is equivalent to the equation for a single fluid with a
vanishing or negligibly small $c_S^2$. The exact condition on eq.(\ref{uperteq})
that we will assume in Fourier space, for a mode with wave vector $\vec{q}$  is
\begin{equation}
 \label{zerocscond}
 (q\eta)^2 c_S^2\ll 1.
\end{equation}
Notice that we restrict ourselves to a positive speed of sound for the dark component. Although a negative speed of sound isn't prohibited these solutions are unstable and restricted in the DE's equation of state parameter space.
The solution of eq.(\ref{uperteq}) is  given by
\begin{equation}
 \label{zerocs}
u(\vec{x},\eta)= C_1(\vec{x}) \theta(\eta)+C_2(\vec{x}) \theta(\eta)\int
d\widetilde\eta \frac{1}{\theta(\widetilde\eta)^2}.
\end{equation}
The first term is the smaller decaying solution and the second is the larger term
which is usually referred to as the ``growing solution" even though it is sometimes
constant, or decays slower than the first term. We are interested only in the
growing solution, because it dominates the solution at late times.

Using eq.(\ref{thetaweq}) we can express the growing solution of eq.(\ref{zerocs}),
\begin{align}
 \label{growing}
\Phi_+(\eta,\vec{x}) = 4\pi G C_2(\vec{x}) \frac{\sqrt{\rho}}{a} \int
d\widetilde\eta a^2(\widetilde\eta)(1+w_{\rm tot}(\widetilde\eta)).
\end{align}

From eq.(\ref{growing}) we can see that the contribution to $\Phi_+$ from an era
when  $w_{\rm tot}=-1$ vanishes. This is to be expected since $w_{\rm tot}$
approaches a constant value of $-1$ only if the DE is a cosmological constant and
it does so at very late times when the cosmological constant completely dominates
the matter.
The solution $\Phi_+(\eta,\vec{x})$ factorizes into a time dependent and spatial part
	\be
		\Phi_+(\eta,\vec{x})=C(\vec{x})\Phi_T(\eta) \\
		C(\vec{x})=4\pi G C_2(\vec{x}) \\
		\Phi_T(\eta)= \frac{\sqrt{\rho}}{a} \int
d\widetilde\eta a^2(\widetilde\eta)(1+w_{\rm tot}(\widetilde\eta))
	\ee
The time dependent part $\Phi_T$ obeys the following differential equation in redshift space
\be
\label{solphi}
 \Phi_T(z)-\frac{2(1+z)}{5+3 w_{\rm tot}(z)} \partial_z\Phi_T(z)& =
\frac{1+w_{\rm tot}(z)}{5+3 w_{\rm tot}(z)}.
\ee
Solving for $w_{\rm tot}$  in terms of $\Phi$ we find
\begin{equation}
\label{ll2}
w_{\rm tot}(z)= -\frac{2(1+z)
\partial_z\Phi_T(z)+1-5\Phi_T(z)}{1-3\Phi_T(z)}.
\end{equation}
The details of the derivation of eqs.(\ref{solphi}) and (\ref{ll2}) can be found in Appendix~\ref{App1}.

The relation between the density perturbation and the metric perturbation is determined (for a spatially flat universe) by the following equation
\be	\label{deltatophi}		 \frac{\delta\rho}{\rho}&=&\frac{2}{3\mathcal{H}^2}\left[\nabla^2\Phi-3{\mathcal{H}}\Phi'- 3{\mathcal{H}}^2\Phi\right].
\ee
For the $\Lambda$CDM model, if one uses the solution for $\Phi_T(z)$ it is possible to show that it is identical to the known solution for the linear density perturbation growth factor which was obtained using different methods \cite{Hu:2000ee}, 
\be
	\Phi_T^{\Lambda\tt{CDM}}(z)&=& \frac{\Omega_{m_0}}{2}(1+z)\frac{H(z)}{H_0}\int\limits_z^\infty dz' (1+z')\left(\frac{H_0}{H(z')}\right)^3.
\ee
The details of the derivation are given in appendix \ref{App2}. Our solution for $\Phi_T(z)$ therefore amounts to a generalization of the known solutions for the linear growth factor to the case of an arbitrary equation of state of the DE.

So far we have not discussed the space dependent factor $C(\vec x)$, which we do
now. The primordial spectrum of $C(\vec x)$ is an input for our analysis. It is
usually assumed to be a power-law spectrum, and that the perturbations are
isotropic and homogeneous.
The primordial spectrum of $\Phi$ can be parametrized as
\be
	P_\Phi(k)=A(k\eta_0)^{n-1}
\ee
The parameter $n$ is the spectral index and $A$ is the spectral amplitude. Both were  measured most recently
by the WMAP experiment \cite{Komatsu:2008hk}. In particular, the spectral index is
approximately $n=.95$ corresponding to an approximately flat spectrum.  The range
over which the spectrum is flat ($n$ is approximately 1) is limited because causal
processes inside the horizon suppress the perturbations \cite{Ma:1995ey}.

The solution for $\Phi$ above is valid only after matter domination. So the primordial spectrum has to be evolved into the ``initial'' spectrum at the beginning of matter domination. We shall use for this purpose the standard practice of including  a transfer function $T_k(\eta)$,
\be
	\Phi(k,\eta)=T_k(\eta)\Phi(k).
\ee
The transfer function is normalized such that $T_k(\eta_0)=1$ for $k\rightarrow 0$. The two-point correlation function  is then given by
\begin{eqnarray}
\label{transf}
	 \langle\Phi(k,\eta)\Phi*(k',\eta')\rangle&=&T_k(\eta)T_{k'}(\eta')k^{-3}P_\Phi(k)(2\pi)^3\delta^3(k-k')
	\nonumber \\ &=& T_k(\eta)T_{k'}(\eta') (2\pi)^3 \left({k\eta_0}\right)^{n-4}  A \delta^3(k-k') .
\end{eqnarray}
To obtain the value of the initial spectrum we must input into eq.(\ref{transf}) the value of $\eta$ at the beginning of matter domination-- $\eta_{in}$. The initial spectrum is then given by
\be
\label{cxofq}
	\langle C(q),C^*(q')\rangle={T_q}^2(\eta_{in})\left(\frac{2\pi}{q^3}\right)^3 \left(\frac{1}{q~\eta_{in}}\right)^{1-n} A\; \delta(q-q').
\ee
The value of ${T_q}^2(\eta_{in})$ can be evaluated using several approximations or calculated numerically using CMBFAST/CAMB etc..
Since during matter domination the perturbations are frozen ($\Phi$ is constant) the exact value of $\eta_{in}$ is not of particular importance.

\section{Expressing $\gamma$ in terms of $\Phi$}
\label{b}

Although it is common to write the shear as a function of the lensing potential we choose to leave it as a function of the metric perturbation using the solution presented in Sec.\ref{a}. The expression for the shear is
\begin{eqnarray}
      \label{gammaphi}
    \gamma=\gamma_1+i\gamma_2&=&
    \int\limits_0^w \!\!dw' \frac{f_K(w-w')f_K(w')}{f_K(w)}
    \left[\frac{\Phi_{|\phi\phi}}{f_K^2(w')\sin^2\theta}-\frac{\Phi_{|\theta\theta}}{f_K^2(w')}
    + 2i\frac{\Phi_{|\theta\phi}}{f_K^2(w')\sin\theta}\right].\ \
\end{eqnarray}
For a single source at distance $w$ the shear is given in eq.(\ref{gammaphi}). We
would like eventually to find the angular moments of the shear-shear two-point
correlation function. Because the shear is not a scalar we have to use the
spin-weight spherical harmonics formalism as in \cite{3dwl3}. The relevant
properties of the $s$-weighted spherical harmonics $\;_{s}Y_{l,m}$ can be found in
\cite{Zaldarriaga:1996xe,3dwl3}. The spin-weight operator $\eth$ operates on the
$s$-weight spherical harmonic and gives an $s+1$-weight spherical harmonic
$\eth\;_s Y_{l,m}=[(l-s)(l+s+1)]^{1/2}\;_{s+1}Y_{l,m}$. Expressed in terms of the
spin-weight operator the shear is given by
 \begin{equation}
 \label{gamma}
  \gamma(w,\theta,\varphi)=\int\limits_0^{w} \!\!dw' \frac{f_K(w-w')}{f_K(w)f_K(w')}
                        \eth \eth \Phi(w',\theta,\varphi).
\end{equation}
For a spatially flat universe $f_K(w)=w$ as can
be seen from eq.(\ref{fk}). Recall in addition that $\Phi$ can be factored into a
space dependent factor $C(\vec{x})$ and time dependent factor $\Phi_T$, defined in
eq.(\ref{PhiT}). Combining these  facts we arrive at the final expression for the
shear,
\begin{equation}
 \label{finalshear}
 \gamma(w,\theta,\varphi)=\int\limits_0^{w} \!\!dw' \frac{w-w'}{w w'}
 \Phi_T(w')   \eth \eth \; C(w',\theta,\varphi).
\end{equation}
Because this is a spin-weight 2 object it can be decomposed into an even and odd parts, these correspond to the well known $E$ and $B$ modes. Using the fact that the $B$ modes of the shear field vanish and to keep things simple we will compute the correlation for the full expression. A detailed explanation of the decomposition and its properties can be found in \cite{3dwl3}.
Now, let us compute the two-point correlation function,
$\langle\gamma(w_1,\theta_1,\varphi_1),\gamma^*(w_2,\theta_2,\varphi_2)\rangle$. We
can use the standard Fourier expansion and the assumption that $\langle
C(\vec{q}_1),C^*(\vec{q}_2)\rangle=f(q)\delta(\vec{q}_1-\vec{q}_2)$  to obtain
\begin{eqnarray}
&&\langle\gamma(w_1,\theta_1,\varphi_1),\gamma^*(w_2,\theta_2,\varphi_2)\rangle =
\cr
   &&= \int\limits_0^{w_1}\!\!du_1  \int\limits_0^{w_2}\!\!du_2 \frac{w_1-u_1}{w_1\;u_1}
   \frac{w_2-u_2}{w_2\;u_2}
    \Phi_T(u_1)\Phi_T(u_2)
    \eth_1\eth_1 \; \bar{\eth}_2\bar{\eth}_2
    \int \!\!d^3 q \:f(q) e^{i\vec{q}\cdot(\vec{x}_1-\vec{x}_2)}.
\end{eqnarray}
We now expand the exponential $e^{i\vec{q}\cdot(\vec{x}_1-\vec{x}_2)}$ in ordinary
spherical harmonics and perform the integration on the unit sphere in $q$-space,
\begin{eqnarray}
&&\langle\gamma(w_1,\theta_1,\varphi_1),\gamma^*(w_2,\theta_2,\varphi_2)\rangle =
\cr &&=(4\pi)^2 \int\limits_0^{w_1}\!\!du_1  \int\limits_0^{w_2}\!\!du_2
\frac{w_1-u_1}{w_1\;u_1}
   \frac{w_2-u_2}{w_2\;u_2}
            \Phi_T(u_1)\Phi_T(u_2)
               \int \!\!dq\: q^2 f(q) \cr
            &&\;\;\;\; \times   \sum_{l,m} j_{l}(qu_1)j_{l}(qu_2)
                       \eth_1\eth_1 \; Y_{lm}(\theta_1,\varphi_1)
                       \bar{\eth}_2\bar{\eth}_2  \;  Y^*_{lm}(\theta_2,\varphi_2).
\end{eqnarray}
The two spin-weight operators act on the ordinary (spin-weight zero) spherical
harmonics in the expansion and give spin-weight $s=2$ spherical harmonics, and
similarly the conjugate spin-weight operators give spin-weight $s=-2$ spherical
harmonics,
\begin{eqnarray}
 &&\langle\gamma(w_1,\theta_1,\varphi_1),\gamma^*(w_2,\theta_2,\varphi_2)\rangle = \cr
 &&=(4\pi)^2 \int\limits_0^{w_1}\!\!du_1  \int\limits_0^{w_2}\!\!du_2 \frac{w_1-u_1}{w_1\;u_1}
   \frac{w_2-u_2}{w_2\;u_2}
 \Phi_T(u_1)\Phi_T(u_2) \int \!\!dq\: q^2 f(q) \cr
 &&\;\;\;\;\times \sum_{l,m} \frac{(l+2)!}{(l-2)!}  j_{l}(qu_1)j_{l}(qu_2)
 \;_2 Y_{lm}(\theta_1,\varphi_1)\;_{-2}Y^*_{lm}(\theta_2,\varphi_2).
\end{eqnarray}
We may perform the summation over $m$ using the summation rule for spin-weight
spherical harmonics
\begin{equation}
 \label{msummation}
\sum_m \;_{s_1}Y_{l,m}^*(\theta_1,\varphi_1)\;_{s_2}Y_{l,m}(\theta_2,\varphi_2)=
\sqrt{\frac{2l+1}{4\pi}}  \;_{s_2}Y_{l,-s_1}(\beta,\alpha) e^{-is_2\delta},
\end{equation}
The angels $\alpha$,$\beta$ and $\delta$ are the rotation angels from
$(\theta_1,\varphi_1)$ to $(\theta_2,\varphi_2)$. The two-point correlation
function should only depend on $\beta$, being the angle between the two directions.
Hence we may choose the polar axis of $\varphi$ such that it is aligned with the
two points and set $\alpha=\delta=0$. In this case, our final expression for the
two-point correlation function is
\begin{eqnarray}
\label{finaltwopoint}
 &&\langle{\gamma}(w_1,\theta_1,\varphi_1),
{\gamma}^*(w_2,\theta_2,\varphi_2)\rangle = \cr &&=(4\pi)^2
\int\limits_0^{w_1}\!\!du_1  \int\limits_0^{w_2}\!\!du_2 \frac{w_1-u_1}{w_1\;u_1}
   \frac{w_2-u_2}{w_2\;u_2}
\Phi_T(u_1)\Phi_T(u_2) \int \!\!dq\: q^2 f(q) \cr &&\;\;\;\;\times \sum_l
\sqrt{\frac{2l+1}{4\pi}}\frac{(l+2)!}{(l-2)!}  j_{l}(qu_1)j_{l}(qu_2) \;_2
Y_{l,2}(\beta,0).
\end{eqnarray}

\subsection*{The shear spin-weight 2 angular power spectrum}

The shear can be expanded in the $s=2$ spin-weight spherical harmonics
$
    \gamma=\sum_{lm}\;_2 a_{lm}\;_2Y_{lm}.
$
From the definition of $\gamma$ in eq.(\ref{gamma}) it follows that $\gamma^*$ is
proportional to $\;_{-2}Y_{lm}$. The conjugation relation of spin-weight spherical
harmonics implies that $ \:_2a_{lm}=\:_{-2}a_{lm}$. From the isotropy and
homogeneity of the shear-shear two-point correlation function we know that it must
be a function of $|\vec{x}-\vec{x}'|$ only, so the two-point function of the
coefficients $\;_2 a_{lm}$ can only depend on $l$,
$
    \langle\;_2 a_{lm},\;_2 a_{l'm'}^*\rangle=\delta_{ll'}\delta_{mm'}\;\;_2 C_l.
$ Consequently, we may express the shear-shear two-point function in terms of the
angular spin-weight two coefficients
 $
    \langle\gamma,\gamma^*\rangle=\sum_l \sqrt{\frac{2l+1}{4\pi}}\;_2C_l\;_2Y_{l,-2}.
$
The summation over $m$ was performed using eq.(\ref{msummation}). By using the
orthogonality relationship of the spin-weight spherical harmonics $ \int d\Omega
\;_s Y_{lm} \;_s Y_{l'm'}^* = \delta_{ll'}\delta_{mm'}$ we can extract the angular
coefficients $\;_2 C_l$ from eq.(\ref{finaltwopoint})
\begin{eqnarray}\label{twoCl}
\;_2C_l  = (4\pi)^2 \frac{(l+2)!}{(l-2)!} \int\limits_0^{w_1}\!\!du_1
\int\limits_0^{w_2}\!\!du_2 \frac{w_1-u_1}{w_1\;u_1} \frac{w_2-u_2}{w_2\;u_2}
\Phi_T(u_1)\Phi_T(u_2)\!  \int\!\! dq\: q^2 f(q) j_{l}(qu_1)j_{l}(qu_2).\ \
\end{eqnarray}

We can also recall now that  for a flat spectrum ($n=1$) the value of $f(q)$ is given in equation (\ref{cxofq}) so that
\be\label{fq}
	f(q)=A\;\left(\frac{2\pi}{q}\right)^3 {T_q}^2(\eta_{in}) .
\ee

The $l$-dependence of the multipoles of the cosmic shear angular power spectrum for the case $n=1$  is shown in Fig.~\ref{fig:spectrum} for three values of redshift. The redshift dependence of the multipoles of the cosmic shear angular power spectrum  is shown in Fig.~\ref{fig:2clofz} for three values of $l$.
\FIGURE[t]{
			\includegraphics{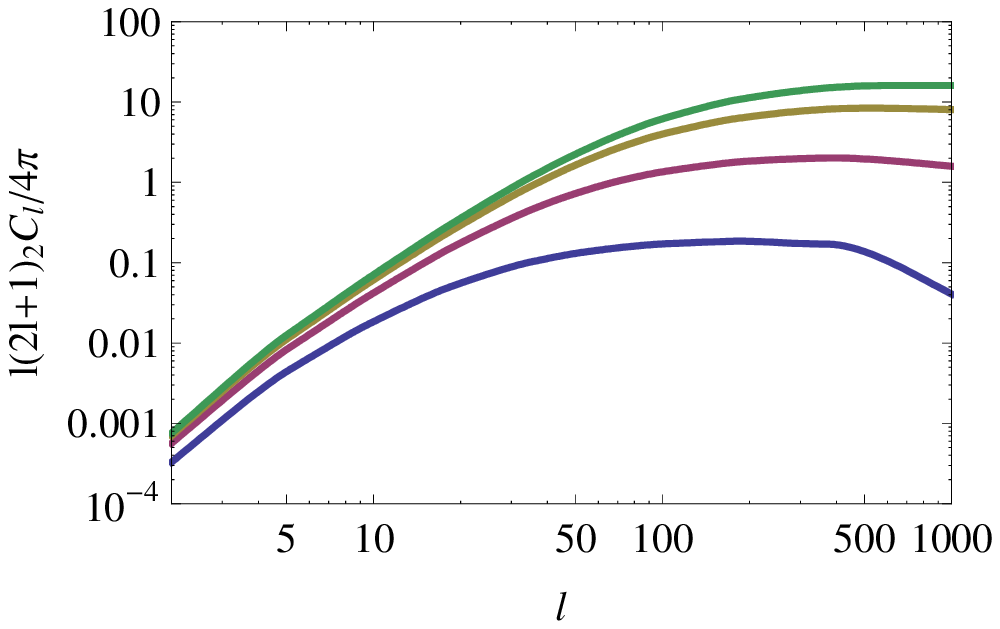}
	\caption{\label{fig:spectrum}The shear spin-weight 2 angular power spectrum for different red shifts. The cosmological model is $\Lambda$CDM with $\{\Omega_{DE}=0.7,\Omega_K=0,n=1,w_{DE}=-1\}$. The four lines correspond to redshifts $z=0.4,1,2,3$ from the bottom up, respectively.}}

The redshift dependence of the multipoles of the cosmic shear angular power spectrum  is shown in Fig.~\ref{fig:2clofz} for three values of $l$.
\FIGURE[t]{
		\includegraphics{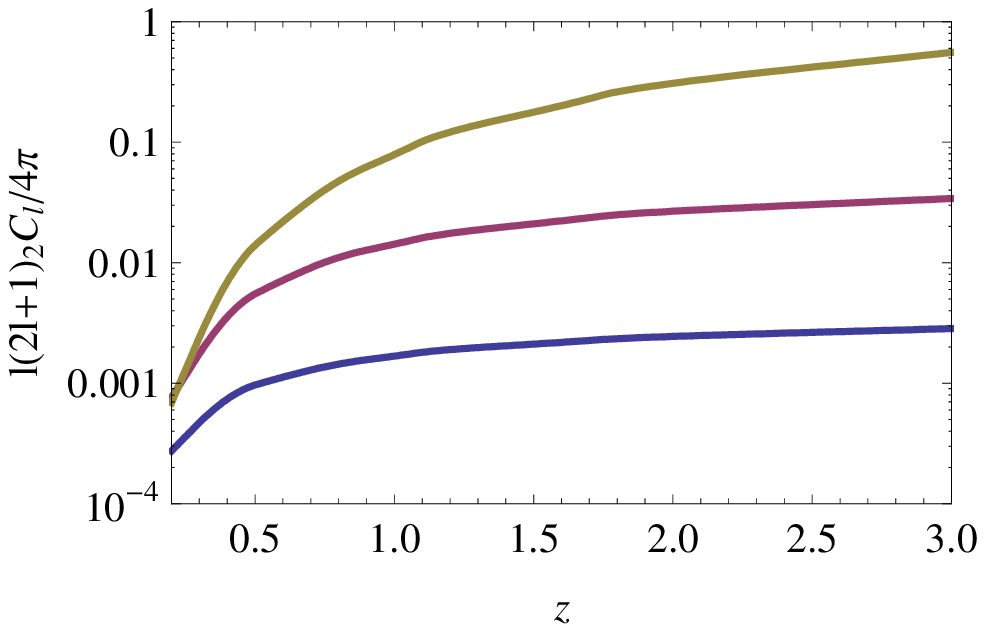}
	\caption{\label{fig:2clofz}The shear spin-weight 2 angular power spectrum for different $l$'s as a function of $z$. The cosmological model is $\Lambda$CDM with $\{\Omega_{DE}=0.7,\Omega_K=0,n=1,w_{DE}=-1\}$. The three lines correspond to Multipols $l=10,30,300$ from the bottom up, respectively.}}
	
The expression for the shear spin-weight 2 angular power spectrum in eq.\ref{twoCl} has a highly oscillatory integrand due to the factor of the spherical Bessel functions. For the case of $w_1=w_2$ we know that for every $q$ there is a cutoff distance such that $qu\approx l$ where the integrand practically vanishes. Since our look back distance is finite to the visible universe, for any given $l$ in the linear part we have a minimal $q$ for which the integrand is still nonzero and below that we won't have any contribution to the power. This property shows up in the angular spectrum as a suppression of the power for the high $l$'s. for such $l$'s $q_{min}$ is already in the region where the power is in the suppressed part of the transfer function. For the redshifts shown in Fig.~\ref{fig:spectrum} this suppression starts above $l=1000$. It can be seen most clearly for the graph for $z=0.4$. In fig.~\ref{fig:2clofz} we can see the effect for all three $l$'s at the very low redshifts. However, a range of red-shifts exists for which the spectrum's evolution is similar for all $l$'s and determined by $\Phi_T$ as we will show next.

\section{red-shift evolution of the cosmic shear angular power spectrum}
\label{c}

In this section we would like to describe and explain some qualitative aspects of the redshift evolution of 3D angular spectrum using our solution for the functional dependence of the shear on $w_{tot}(z)$. We will show that the $_2C_l(z)$'s depend weakly  on the shape of the initial spectrum (the spectrum at the beginning of matter domination) and more importantly that the redshift evolution is quite  sensitive to changes in $w_{tot}$ independently of the initial spectrum. Since we wish to  focus on determining the qualitative aspects of the spectrum's redshift evolution, we will use approximations that are simple enough to allow us to obtain  analytical expressions.
Of course, the spectrum can be evaluated numerically  very accurately for each specific cosmology using CMBFAST/CAMB etc.. using the same techniques that are applied to the CMB.
In addition to the qualitative analysis we will show accurate numerical solutions for the case of $\Lambda$CDM (using CAMB) to supplement our qualitative results.

In the above and subsequent calculations and analysis several assumptions are made for the sake of simplicity. The validity of these assumptions is generally accepted. However, the sensitivity of our results to them should be explored further either analytically or numerically. One such assumption that may be particularly relevant is spatial flatness. We hope to discuss it in future work.

From the epoch of cold matter dominance onward, the time evolution of the perturbation $\Phi_k$ is independent of $k$ (under the assumptions specified previously). From the solution eqs.(\ref{growing})-(\ref{solphi}) one can see that the time dependent part $\Phi_T(z)$ has a simple and known functional dependence on $w_{tot}$. It follows that if we had  full knowledge of the 3D spectrum at different times we could extract $w_{tot}$. It was shown that the 3D spectrum in $k-$space can be used to constrain the cosmological parameters \cite{Heavens:2006uk,Kitching:2006mq}.  We have found that $w_{tot}$ (and thus the expansion history of the universe) can also be constrained from the redshift evolution of angular power spectrum multipoles. However, a subtle effect complicates matters: Looking farther in the radial direction necessarily involves looking back in time. This effect forces a mixing of the spatial and time dependent parts of the perturbation.

The 3D spectrum $P_\gamma(k)$ evolves in time independently of $k$. Its multipole expansion in terms of the multipoles $_2C_l$ is, however, distance-dependent and therefore the redshift evolution of the different multipoles becomes $k$-dependent. We will show that this dependence on $k$ is advantageous and useful.  In the expression for the multipoles both $\Phi_T(z)$ and the luminosity distance $d_L(z)$ appear. Both quantities depend on $w_{tot}(z)$ in a different way making them more sensitive to changes in $w_{tot}$ at different redshifts. More precisely, their sensitivity to changes in $w_{tot}$ varies in opposite ways. We have found that due to this, in the expression for the multipoles, their explicit combination is less degenerate in $w_{tot}$ than each of them separately.

\subsection{The qualitative dependence of the spectrum on $\Phi_T$}

To understand how the $_2C_l$'s depend on $\Phi_T(z)$ we will assume for the moment  a flat (constant) initial spectrum.  We discuss a flat spectrum because it gives us better insight to the qualitative behavior of the solutions. We will then relax this oversimplification.
A flat initial spectrum corresponds to assuming a flat primordial spectrum ($n=1$) and $T_q(z_{in})=1$.  The solution in this case can be obtained analytically in a closed form. Moreover, with a constant transfer function in eqs.~(\ref{twoCl}-\ref{fq}) it becomes possible to calculate the angular moments for any spectral index in terms of hypergeometric functions.
For a flat spectrum ($n=1$) the result is extremely simple
\begin{equation}
 \label{qint}
 \int \!\!dq \frac{1}{q} j_{l}(qu_1)j_{l}(qu_2)= \frac{1}{2 l }
 \begin{cases}
   \left(\frac{u_1}{u_2}\right)^{l}  & u_2> u_1   \cr
 \left(\frac{u_2}{u_1}\right)^{l} & u_2 <  u_1.
\end{cases}
\end{equation}
The function in eq.(\ref{qint}) is essentially a delta function,
\begin{equation}
\label{approxqint}
\begin{cases}
   \left(\frac{u_1}{u_2}\right)^{l}  & u_2> u_1   \cr
 \left(\frac{u_2}{u_1}\right)^{l} & u_2 <  u_1
\end{cases}
\simeq  \frac{2 l}{l^2-1} u_1 \delta(u_1-u_2),
\end{equation}
whose normalization  is determined by the integral
 $ \int_0^{u_1} du_2 \left(\frac{u_2}{u_1}\right)^{l} +
  \int_{u_1}^{\infty} du_2   \left(\frac{u_1}{u_2}\right)^{l}
= \frac{u_1}{l^2-1}$. The approximation in eq.(\ref{approxqint}) becomes better for
larger $l$'s, exactly the range of $l$'s of interest.
Putting everything together we obtain,
\begin{eqnarray}
  \;_2C_l &=&
   (4\pi)^2 A \frac{(l+2)!}{(l-2)!} \int\limits_0^{w_1}\!\!du_1
\int\limits_0^{w_2}\!\!du_2 \frac{w_1-u_1}{w_1\;u_1} \frac{w_2-u_2}{w_2\;u_2}
\Phi_T(u_1)\Phi_T(u_2) \frac{1}{(l^2-1)} u_1 \delta(u_1-u_2) \cr &=& (4\pi)^2 A
l(l+2) \int\limits_0^{min[w_1,w_2]}\!\!\!\!\frac{du}{u}
\frac{(w_1-u)(w_2-u)}{w_1\;w_2} \Phi_T^2(u).
\end{eqnarray}
We can assume without loss of generality that $w_1<w_2$, so our final result for this case is
\begin{equation}
\label{final2cl}
 \;_2C_l(w_1,w_2)=(4\pi)^2 A l(l+2) \int\limits_0^{w_1}\frac{du}{u}
\frac{(w_1-u)(w_2-u)}{w_1\;w_2} \Phi_T^2(u)\;\;,\;\;w_1<w_2.
\end{equation}
We observe that the $_2C_l$'s depend on $w_{tot}$ through an integral expression of the square of $\Phi_T$ and a kernel function of $u=\frac{d_L(z)}{1+z}$. We recall that eq.(\ref{final2cl}) is derived assuming a constant transfer function.
Since $\Phi_T(0)$ is finite, the integral in eq.(\ref{final2cl}) is formally
divergent. The formal divergence at small $u$ is not physical rather it is a property of blue or flat spectra
($n\ge 1$). In technical terms, tracing back the properties of the small $u$ region
of the integrand, one sees that it corresponds to the region of high $q$'s in the
integrand of eq.(\ref{qint}).
To further understand the behavior of our integral and the influence of the transfer function on the result let us look at the evolution of perturbations before the epoch of matter domination.
During matter domination $\Phi$ is constant but during radiation domination the solutions for $\Phi$ inside the horizon decay as $1/a^2\sim 1/\eta^2$. For values of $q$ larger than some maximal value $q_{\rm max}$ the solutions of $\Phi$ are therefore completely suppressed, and thus the $u$ integral is effectively cutoff at $u\propto 1/q_{\rm max}$ and becomes finite.

The real transfer function undergoes a smooth transition from unity at small $q$ to zero at high $q$'s (it should be evaluated  in the transition region by approximations such as BBKS). If we use accurate approximations of the transfer functions shape in Eq.(\ref{twoCl}) the analytic solutions become too complicated and it is hard to gain any understanding from it. Rather, we will demonstrate the effect of the suppression of high $q$'s using a step function approximation of the transfer function, representing sharp cutoff on the spectrum at some $\widetilde{q}=q_{\rm max}$.

When the transfer function can be approximated by a step function $T_q=1-\theta(q-\widetilde{q})$ then the integral in eq.~(\ref{solphi})  becomes
\be
	\int\limits_0^{\infty} \frac{dq}{q}j_l(u_1 q)j_l(u_2 q)(1-\theta(q-\widetilde{q}))=
		\int\limits_0^{\widetilde{q}} \frac{dq}{q} {j_l}^2(q u_1) \frac{1}{l^2-1}\frac{u_1}{l+1}\delta(u_1-u_2).
\ee
The definite integral
$
2l(l+1) \int\limits_0^{\widetilde{x}} \frac{dx}{x} {j_l}^2(x)
$
depends for large $l$ on $x/l$ so it can be approximated (numerically) for large $l$'s and for $(\tilde{x}/l)>1$ as
\be
\label{lfunc}
2l(l+1) \int\limits_0^{\tilde{x}} \frac{dx}{x} {j_l}^2(x)\approx \left(\frac{\ln(\tilde{x}/l)}{\ln a}\right)^{b}.
\ee
where $a,b$ are some $l$-independent constants. The integral vanishes for $(\widetilde{x}/l)<1$.
The expression for the shear becomes
\be
\label{stepFinal}
&&\;_2C_l = 4A(2\pi)^5 l(l+2)
\int\limits_{l/\tilde{q}}^{\min[w_1,w_2]} \left(\frac{w_1-u}{w_1}\right)
\left(\frac{w_2-u}{w_2}\right)  \left(\frac{\ln(\tilde{q} u/l)}{\ln a}\right)^{b}{\Phi_T}^2(u) \frac{d u}{u}.
\ee
The resulting expression (\ref{stepFinal}) is different from the one in eq.(\ref{final2cl}) in that its integrand has an added $l$-dependent numerical factor (\ref{lfunc}) and, more importantly, it is finite. The functional dependence on $w_{tot}$ of both expressions comes from $\Phi_T$, so this dependence is shared whether one assumes an unrealistic trivial transfer function ($T_q=1$) or a more realistic step function form to the transfer function. Based on this comparison,  we expect that the functional dependence of the solution with the correct, fully realistic and numerically calculated transfer function will also have the same simple dependence on $w_{tot}$ that comes from $\Phi_T$.

\subsection{The sensitivity of the $_2C_l$'s evolution with red-shift to changes in $w_{tot}$}

\FIGURE[t]{
		\includegraphics[height=5cm]{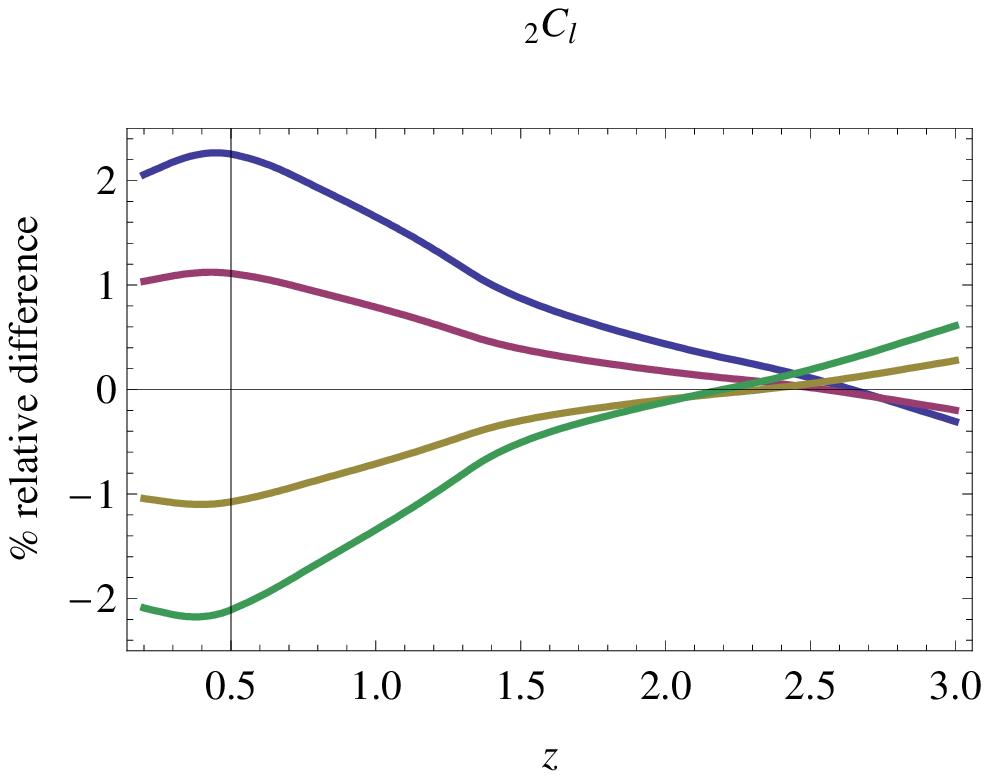}
		\includegraphics[height=5cm]{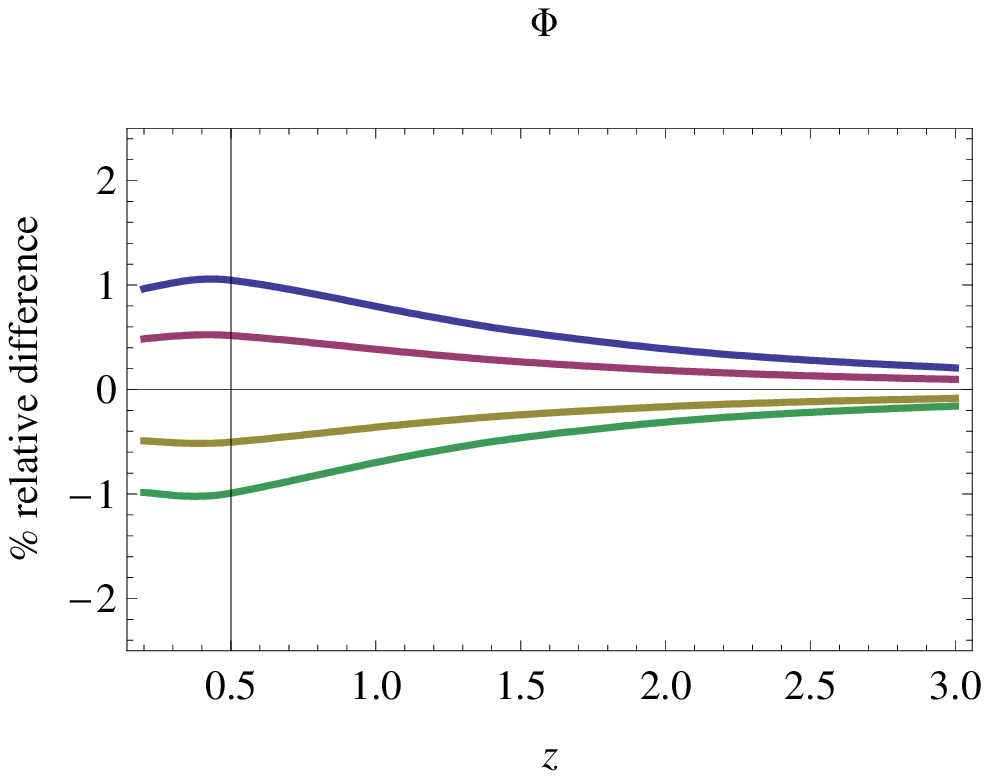}
		\includegraphics[height=5cm]{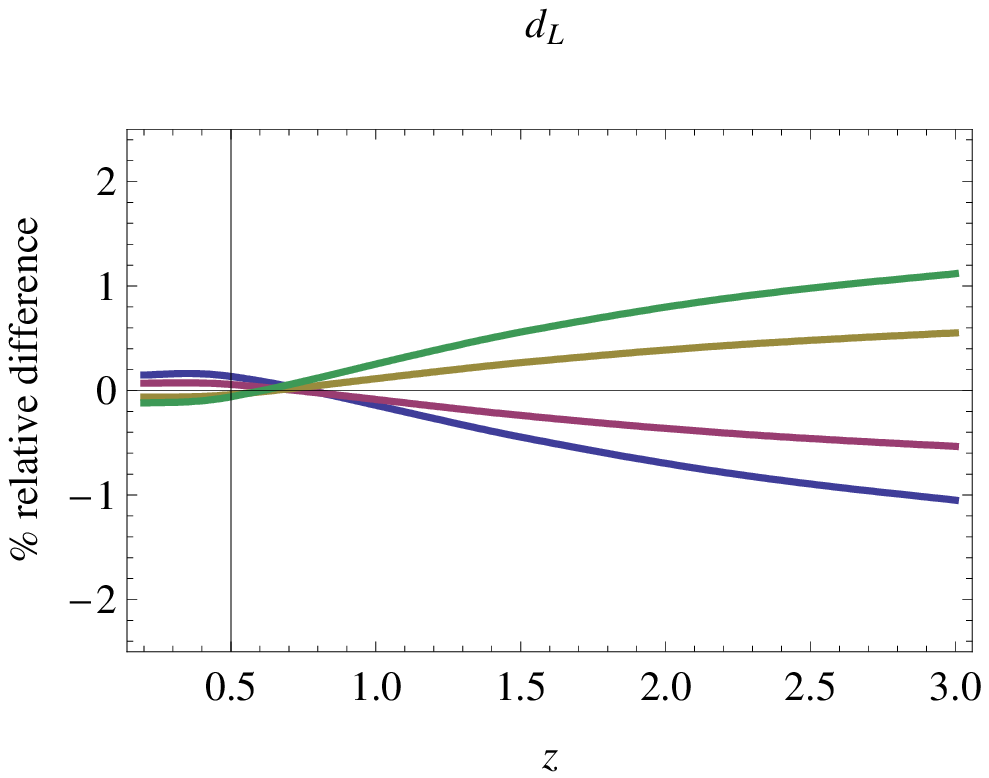}
	\caption{\label{fig:diff}The three panels show the relative difference (in percent) as function of red-shift between different cosmological models and a $\Lambda$CDM concordance model. The four lines correspond to the four sets of $(\Omega_{DE},w_{DE})$ parameters: $(0.72,-0.95),(0.71,-0.975),(0.69,-1.025),(0.68,-1.05)$. The first panel shows the relative difference of the $\;_2 C_l$'s for $l=100$, the second is of $\Phi_T$ and the third shows $d_L$.  $\Phi_T$ and $d_L$ differ in their sensitivity.}}

In the previous subsection we have determined the dependence of the angular power spectrum on $\Phi_T$. But, it is the combination of the evolution of the perturbations convoluted with the geometrical kernel that determines the true sensitivity of the angular spectrum to changes in $w_{tot}$. In Fig.~\ref{fig:diff} we show the $\;_2C_l$'s for four different cosmological models. Also shown is the luminosity distance $d_L$, and $\Phi_T$ for each of the models. It is evident from the figure that the $\Phi_T$'s are more sensitive to changes in the expansion history at low redshifts while the $d_L(z)$'s are more sensitive to such changes at higher redshifts. It is also evident that $\;_2C_l$'s sensitivity follows $\Phi_T$, so we may argue that the $\;_2C_l$'s ``measure" $\Phi_T$. The sensitivity of the  $\;_2C_l$'s is twice as much as that of $\Phi_T$ simply because they depend on $(\Phi_T)^2$. This means that to distinguish between these models using luminosity distance measurements we would need  accuracies at the sub-percent level, while they could be easily distinguished if the $\;_2C_l$'s can be measured with about a percent accuracy.
The realistic prospects of constraining the total equation of state and other cosmological functions depends on how well it will be possible to handle the low redshift range and on the selections that are made in the survey and in the data analysis. Hence, proper numerical analysis of the power spectrum's sensitivity $\frac{\delta\;_2C_l(z)}{\delta w_{tot}}$ and the likelihood calculations are necessary. These are left for  future work.

\section{Conclusions and outlook}
\label{e}
Several studies of the ability to use 3D weak lensing measurements to constrain cosmological parameters exist in the literature. To the best of our knowledge, all these studies have presented the numerical results for specific fiducial lensing surveys with weakly coupled DE that has an equation of state of the form $w_{DE}=w_0+w_1 z/(1+z)$ and positive speed of sound. These studies are important in demonstrating the advantages of three dimensional analysis over the two dimensional one. However, it is not easy to extract from them information on the functional dependence of the 3D spectrum on the equation of state and hence on the degeneracies with respect to simultaneous changes in several cosmological parameters. Hence, it is harder to use them to forecast the possible improvements in the constraints that one should get from more accurate shear measurements.

By calculating the functional dependence of quantities such as the 3D cosmic shear on $w_{tot}$ in a fully relativistic and without assuming a specific functional form for $w_{tot}$, we can estimate the theoretical limits on the constraints from a given accuracy for shear measurements. Our conclusion is that the red shift evolution of the spin weight two angular power spectrum of shear is the most sensitive measure. In this paper we have presented our analytical results and some numerical examples for the simple case of Linder's parameterization thus laying the ground for a more detailed, application oriented, analysis.
We have found that redshift evolution of the angular power spectrum of the three dimensional cosmic shear can
be used to determine the expansion history of the universe. This method is
sensitive to other prior assumptions about the functional dependence of $w_{\rm
tot}$ on redshift than kinematic methods that rely on the homogeneous and isotropic
universe (such as luminosity distance etc.).

The implementation of our method will require extraordinary observational efforts
because it requires very accurate measurements of the shear. Further, good
determination of the three dimensional cosmic shear requires
accurate determination of redshifts of the sources, and in addition a good
measurement of luminosity distance (or angular distance) as a function of redshift.
To achieve such ambitious observational goals would probably require combining the
results of several accurate experiments.

The equations for the metric perturbation $\Phi$, unlike the equations for the
density perturbation $\delta\rho$,  can be adapted in a straightforward manner to
other metric theories of gravity which modify Einstein's general relativity. The
equations can be easily modified to account for various corrections, and hence our
methods can be used,  in principle, to test modifications to general relativity. A
discussion of this point can be found in \cite{Bertschinger:2006aw}.

\section*{Acknowledgments}

We are grateful to Ed Bertschinger for sharing with us his results
\cite{Bertschinger:2006aw} a long time before their publication and for pointing
out their relevance to the use of perturbation probes for determining the expansion
history of the universe. We thank D. Eichler, S. Hofmann, G. Kane,  I. Maor, A.
Nusser, M. Perry and A. Zytkow for useful discussions and comments.




\appendix
\section{Expressing $w_{\rm tot}$  in terms of $\Phi$}\label{App1}

We would like to transform eq.(\ref{growing}) into a more convenient form. First,
we change the integration variable from $\eta$ to $\rho$ by using the conservation
equation
\begin{equation}
\label{dlnrho}
 d\ln\rho=-3(1+w)d\ln a.
\end{equation}
Since $d\ln a/d\widetilde\eta=\H(\widetilde\eta)$, it follows that
$d\widetilde\eta=d\ln a \frac{d\widetilde\eta}{d\ln a}= -1/3
\frac{d\rho}{\H(\rho+p)}$, then the growing solution is given by
\begin{equation}
 \label{growingrho}
\Phi_+(\eta,\vec{x}) = 4\pi G C_2(\vec{x}) \frac{\sqrt{\rho}}{a}\int \frac{1}{3}
a^2 \frac{d\rho}{\H\rho}
\end{equation}
(the sign was absorbed in changing the integration boundaries.)

If the universe is spatially flat then from Friedman's equation it follows
that $\H^2=\frac{8\pi G}{3} a^2 \rho$. In this case eq.(\ref{growingrho})
simplifies,
\begin{equation}
 \Phi_+(\eta,\vec{x}) = \sqrt{\frac{2 \pi G}{3}} C_2(\vec{x})\frac{\sqrt{\rho}}{a} \int
d\rho\frac{a(\rho) }{\rho^{3/2}}.
\end{equation}
Because the spatially flat case is simpler we will assume a spatially flat universe
for the rest of this discussion.

We may further change variables to
\begin{equation}
\label{defchi}
 \chi=\frac{a/a_0}{\sqrt{\rho/\rho_0}}.
\end{equation}
Here $a_0$ and $\rho_0$ are the values of the scale factor and the energy density
today. The value of the variable $\chi$ today is $\chi_0=1$ and it vanishes for
very early times, if the universe was matter dominated as expected. Consequently,
the range of $\chi$ is $0\le\chi \le 1$.

Using eq.(\ref{dlnrho}) we get $d\ln \chi= d\ln a-1/2 d\ln \rho$, so $d\ln
\rho=-\frac{6(1+w_{\rm tot})}{3(1+w_{\rm tot})+2} d\ln \chi$ and therefore
$d\rho\frac{a(\rho) }{\rho^{3/2}}= \chi= d\ln\rho  -\frac{6(1+w_{\rm
tot})}{3(1+w_{\rm tot})+2} d\chi$ so we finally get
\begin{align}
\label{phichi}
 \Phi_+(\chi,\vec{x})& = \sqrt{24 \pi G} C_2(\vec{x})\frac{1}{\chi} \int\limits_0^\chi
 \!\!d\widetilde{\chi}
\frac{1+w_{\rm tot}(\widetilde{\chi})}{5+3 w_{\rm tot}(\widetilde{\chi})}.
\end{align}

The solution $\Phi_+(\chi,\vec{x})$ factorizes
\begin{align}
\label{factorphiT}
 \Phi_+(\chi,\vec{x}) & = C(\vec{x})\Phi_T(\chi)  \\
C(\vec{x})& = \sqrt{24 \pi G} C_2(\vec{x}) \\
 \label{PhiT}
\Phi_T(\chi) & =\frac{1}{\chi} \int\limits_0^\chi
 \!\!d\widetilde{\chi}
\frac{1+w_{\rm tot}(\widetilde{\chi})}{5+3 w_{\rm tot}(\widetilde{\chi})}.
\end{align}

We can invert eq.(\ref{PhiT}), since $\partial_\chi
\left(\chi\Phi_T\right)=\frac{1+w_{\rm tot}(\chi)}{5+3 w_{\rm tot}(\chi)}$. It
follows that
\begin{align}
\label{chiphi}
 w_{\rm tot}(\chi)& = -
\frac{1- 5 \partial_\chi \left(\chi\Phi_T\right)}{1-3\partial_\chi
\left(\chi\Phi_T\right)}.
\end{align}

If we so wish we can also express $\Phi_T$ and $w_{\rm tot}$ as a function of
redshift $z=a_0/a-1$. Since $\chi=\frac{a}{\sqrt{\rho}}$, ${\chi}=0$ corresponds to
$z\to\infty$, and ${\chi}=1$ corresponds to $z=0$. For a spatially flat universe,
as we are considering
\begin{equation}
\label{defchih}
{\chi}=\frac{a/a_0}{\sqrt{\rho/\rho_0}}=\frac{1}{1+z}\frac{1}{H(z)/H_0}.
\end{equation}
From eq.(\ref{PhiT}) we find
\begin{align}
\label{phiz1}
 \Phi_T(z)& = \frac{1}{\widetilde{\chi}(z)}
  \int\limits_\infty^z \!\!d\widetilde{z}\ \frac{d\widetilde{\chi}(\widetilde{z})}{d\widetilde{z}}
\frac{1+w_{\rm tot}(\widetilde{z})}{5+3 w_{\rm tot}(\widetilde{z})},
\end{align}
or equivalently, a differential equation for $\Phi_T$
\begin{align}
\label{phiz4}
 \Phi_T(z)+\frac{{\chi}(z)}{\partial_z {\chi}(z)} \partial_z\Phi_T(z)& =
\frac{1+w_{\rm tot}(z)}{5+3 w_{\rm tot}(z)}.
\end{align}
From eq.(\ref{defchi}) it follows that
$
\partial_z\ln
{\chi}(z)=\partial_z\ln(a/a_0)-\frac{1}{2}\partial_z\ln\rho,
$
so
$\frac{\partial_z{\chi}(z)} {{\chi}(z)}=\partial_z\ln(\frac{1}{1+z})+
\frac{3}{2}(1+w_{\rm tot})\partial_z\ln(\frac{1}{1+z}), $ and the final result is
that
\begin{equation}
 \label{dchiz}
\frac{\partial_z{\chi}(z)} {{\chi}(z)} =-\frac{1}{1+z}-\frac{3}{2} \frac{1+w_{\rm
tot}(z)}{1+z}=- \frac{5+3 w_{\rm tot}(z)}{2(1+z)}.
\end{equation}
Substituting eq.(\ref{dchiz}) into eq.(\ref{phiz4}) we get
\begin{align}
\label{phiz6}
 \Phi_T(z)-\frac{2(1+z)}{5+3 w_{\rm tot}(z)} \partial_z\Phi_T(z)& =
\frac{1+w_{\rm tot}(z)}{5+3 w_{\rm tot}(z)}.
\end{align}
Equation ({\ref{phiz6}) allows us to solve for $\Phi_T(z)$ in terms of $w_{\rm
tot}$. We may define an integration factor
\begin{equation}
I(z)=\int\limits_z^{z_{\rm in}} \!\!d\widetilde{z}\ \frac{5+3 w_{\rm
tot}(\widetilde{z})}{2(1+z)},
\end{equation}
that simplifies eq.({\ref{phiz6})
\begin{equation}
e^{-I(z)}\partial_z\left[e^{I(z)} \Phi_T(z)\right]= -\frac{1+w_{\rm
tot}(z)}{2(1+z)}.
\end{equation}
The initial conditions on $\Phi_T$ are prescribed at the initial redshift $z_{\rm
in}$. In terms of $I(z)$,
\begin{equation}
\Phi_T(z)=\Phi_T(z_{\rm in})+ e^{-I(z)} \int\limits_z^{z_{\rm in}}
\!\!d\widetilde{z}\ e^{ I(\widetilde{z})} \ \frac{1+w_{\rm
tot}(\widetilde{z})}{2(1+\widetilde{z})}.
\end{equation}
Here we have integrated eq.(\ref{phiz6}) from large redshifts towards smaller ones.
We can also integrate eq.(\ref{phiz6}) from small redshifts towards larger ones.
This will require knowing the amplitude of the perturbation at late times which is
harder to determine.

From eq.({\ref{phiz6}) we can also solve for $w_{\rm tot}$ in terms of $\Phi_T$,
\begin{align}
 \label{wtotalz}
w_{\rm tot}(z)= -\frac{2(1+z)
\partial_z\Phi_T(z)+1-5\Phi_T(z)}{1-3\Phi_T(z)}.
\end{align}


\section{The relation between $\Phi_T(z)$ and the linear growth factor}\label{App2}

For the $\Lambda$CDM model there is an exact solution for the evolution of the matter density perturbations in the linear regime. The growth rate $F(z)/(1+z)$ for the linear density perturbations $\delta_m(z)=\delta(0)F(z)/(1+z)$ is \cite{Hu:2000ee}
\be\label{hu}
	F(z)\propto (1+z)\frac{H(z)}{H_0}\int\limits_z^\infty dz' (1+z')\left(\frac{H_0}{H(z')}\right)^3.
\ee
To find the relation between $F(z)$ and our solution for $\Phi_T(z)$ we use eq.(\ref{deltatophi}) in the Newtonian approximation
\be
	\frac{\delta\rho}{\rho}=\frac{2}{3(a H)^2}\nabla^2\Phi
\ee
so that
\be
	\delta_m(z)&=&\frac{\delta\rho_m}{\rho_m}
											=\frac{\delta\rho}{\rho}\frac{\rho}{\rho_m}
						=\frac{2}{3{H_0}^2}\frac{\Phi_T(z)}{\Omega_{m_0} (1+z)}.
\ee
Thus $F(z)\propto \Phi_T(z)/\Omega_{m_0}$. 

To show that the solutions are indeed identical we start from the solution for the metric perturbation. The growing mode solution of eq.~(\ref{growing}) in cosmic time is given by
\be
	\Phi_T(t)=\frac{\sqrt{\rho}}{a}\int\limits^t d\tilde{t} a(\tilde{t})(1+w_{tot}(\tilde{t})).
\ee
Using the fact that $H^2=\frac{8\pi G}{3}\rho$ and $H dt=-dz/(1+z)$ we express $\Phi_T$ as a function of $z$
\be
	\Phi_T(z)&\propto&(1+z)H(z)\int\limits_z^\infty \frac{dz'}{(1+z)^2 H(z')}(1+w_{tot}(z'))\cr
						&=&(1+z)H(z)\int\limits_z^\infty \frac{dz'}{H^3(z')}\frac{1+z'}{(1+z')^3}(\rho+p).
\ee
For$\Lambda$CDM, $w_{DE}=-1$ and therefore $\rho+p=\rho_m$. Using the fact that $\rho_m={\rho_m}(0)(1+z)^3$ we finally get
\be
\label{solhu}
	\Phi_T(z)	&\propto&\Omega_{m_0}(1+z)\frac{H(z)}{H_0}\int\limits_z^\infty dz' (1+z')\left(\frac{H_0}{H(z')}\right)^3,
\ee
where $\Omega_{m_0}=\Omega_{m}(0)$.
The solution in eq.~(\ref{solhu}) is identical to the one appearing in eq.~(\ref{hu}). The growth factor in eq.~(\ref{hu}) is normalized such that $F(0)=1$. The normalization that we choose in the paper following \cite{mukhanov} is such that $\Phi_T(z\gg 1)=1/5$ outside the horizon and thus numerically
\be
	\Phi_T=\frac{{\Omega_{m_0}}}{2}(1+z)\frac{H(z)}{H_0}\int\limits^\infty_z dz'(1+z')\left(\frac{H_0}{H(z')}\right)^3.
\ee
The full amplitude of the metric perturbation is related to the density perturbations amplitude through the same relation as in \cite{Hu:2000ee}.

\end{document}